# Finite Fault Analysis and Near-Field Dynamic Strain and Rotation Estimates due to the 11/05/2011 (Mw5.2) Lorca Earthquake, South-Eastern Spain.

By Miguel A. Santoyo[1]

[1]. Departamento de Geofísica y Meteorología, Facultad de Ciencias Físicas, Universidad Complutense de Madrid. Plaza de Ciencias s/n, Ciudad Universitaria, 28040, Madrid, Spain.  e-mail: masantoyo@pdi.ucm.es

**ABSTRACT**

The 11/5/2011 Lorca, Spain earthquake (Mw=5.2) and related seismicity produced extensive damage in the town of Lorca and vicinity. During these earthquakes, evidence of rotations and permanent deformations in structures were observed. To analyze these aspects and study the source properties from the near–field, the displacement time histories were obtained including the static component at Lorca station. Displacement time histories were computed by an appropriate double time integration procedure of accelerograms. Using these data, the foreshock and mainshock slip distributions were calculated by means of a complete waveform kinematic inversion. To study the dynamic deformations, the 3D tensor of displacement gradients at Lorca station was first estimated by a single station method. Using the finite fault inversion results and by means of a first order finite difference approach, the dynamic deformations tensor at surface was calculated at the recording site. In order to estimate the distribution of the peak dynamic deformations, the calculation was extended to the close neighboring area of the town. The possible influence of the near-field deformations on the surface structures was analyzed.

## INTRODUCTION

The 11/5/2011 Lorca, Spain earthquake (Mw=5.2) showed with severity the damaging potential of relatively small shallow earthquakes rupturing near populated areas (*e.g. 1960 Agadir, Morocco, Mw 5.7; 1969 Yangjiang, China, Mw 5.9; 1986 El Salvador, Mw5.5).* During this seismic series, nine people died and about 300 people were injured. Extensive damages to both historical and recently constructed buildings were produced within the urban area, including several partial and total collapses of buildings and churches (Feriche 2011; Giner-Robles et al. 2011; IGN 2011). Among others, strong structural damages on buildings like large permanent deformations of ground floor storeys, shear failures on columns and cracks on bearing walls were extensively observed (e.g. Vidal et al. 2011; Murphy 2011). Possible geotechnical failures were also recognized (Vidal et al. 2011). Non-structural damages included the extensive falling of parapets and external masonry panels. A special attention was given to the observational evidence of rotational effects on structures. Examples of this were the permanent rotations observed after the earthquakes on some blocks of an obelisk, and the rotation of spires or pinnacles at the top of some church towers (Feriche 2011; Vidal et al. 2011). Giner-Robles et al. (2011) reported strong permanent deformations on historical structures including torsions on several church bell-towers.

Some damaging effects produced by earthquakes may be the result of high dynamic deformations (strains and rotations) rather than from peak accelerations or other displacement variations (e.g. Clough and Penzien 1993; Bodin et al. 1997; Stupazzini et al. 2009, Smerzini et al. 2009). An example of this were the damages produced by the January 17, 1994 Nothridge, California earthquake (Mw 6.7), where non-linear soil response occurred at many locations in the San Fernado Valley (e.g. Todorovska and Trifunac 1996). Surface dynamic deformations induced by earthquakes have been studied in the past for different regions (e.g. Spudich et al. 1995; Bodin et al. 1997; Gomberg 1997; Paolucci and Smerzini 2008). Pioneering works (e.g. Bouchon and Aki 1982; Lee and Trifunac 1985; Lee and Trifunac 1987) have showed that rotational ground motions can be important in the near-field. After them, the rotational effects induced by earthquakes have been gaining interest among the scientific community (e.g. Lee et al. 2009). As recent examples,

Stupazzini et al. (2009) studied the rotational ground motion effects in the near-field region for the Grenoble Valley and Cucci and Tertuliani (2011) studied the rotational effects due to the L'Aquila 2009 earthquake in Italy.

In this work, the finite rupture slip distribution of the 11/5/2011 Lorca Mw=4.6 foreshock and Mw=5.2 mainshock, were obtained by linear kinematic inversion of the complete waveforms. After this, the 3D tensor of dynamic deformations was computed from data by the aid of a single station method, assuming the incidence of body waves through the closest recording site. Using the obtained finite fault results and by means of a first order finite difference approach for the computation of the spatial derivatives of ground motion, it was calculated the tensor of synthetic dynamic deformations at surface. After comparing the results from both methods, calculations were extended to the neighboring area of the town, in order to estimate the peak dynamic deformation distribution at surface. The possible influence of these free-field deformations in the urban area was analyzed.

*Tectonic setting and recent seismicity.*
Seismicity in southern Spain is mainly governed by the convergence between the African plate and the Eurasian plate and characterized by low to moderate earthquake magnitudes. In the western Mediterranean, the relative velocity between the African plate with respect to the Eurasian plate varies between 4 and 9 mm year$^{-1}$ (e.g. Nocquet and Calais 2003).

The Murcia region is located in the eastern side of the Betics Cordillera in the western part of the European Alpine chain. The Betics are divided in the Internal and the External zones, where the Sierra España area comprises the contact between the Alpujárride-Maláguide metamorphic complexes with different Mesozoic and Cenozoic sedimentary materials. This contact is characterized by different fault systems. One of the main systems in this boundary strip is the Alhama de Murcia Fault (e.g. Martinez-Díaz 2002). Figure 1 shows the location map and tectonic setting of the studied area. In this Figure, CRF =Crevillente fault; NBF =North-Betic fault; AMF=Alhama de Murcia fault. The principal directions of its surface expressions are oriented between the azimuths 40° and 80° (Martinez-Díaz 2002). Different fault systems including the AMF have been associated with the seismicity in this zone, characterized by focal mechanisms ranging from pure strike slip to steep thrusting at shallow depths (e.g. Buforn et al. 1995). Since the middle of the past century, several moderate earthquakes (M>5) have occurred in this region. Between 1995 and 2010, four seismic series have taken place in this zone with main shocks in Mula (1995, Mw 4.1 and 1999, Mw 4.9), Bullas (2002, Mw 5.0) and Bullas-La Paca (2005, Mw 4.7) (e.g. Santoyo and Luzón 2008).

The 11/5/2011 (Mw5.2) Lorca earthquake series have been attributed to the AMF system activity (Martinez-Díaz and Alvarez-Gómez 2011; Vissers and Meijninger 2011). The foreshock and mainshock epicenters were located 5.0 km and 6.0 km N-NE the town of Lorca. The foreshock, mainshock and major aftershock depths were determined between 1.0 km and 6.0 km (Lopez-Comino et al 2012). At least, 47 regional broad-band and 13 accelerometric stations recorded the three main events of the series. From them, the closest recording site was the Lorca accelerographic station, located few kilometres from the epicenter. Due to its proximity to the ruptured fault, relatively high peak horizontal accelerations (0.367g) were observed during the mainshock. Lopez-Comino et al. (2012) relocated the series including the three major events, analyzed their focal mechanisms and obtained the rupture directivity of the mainshock. They obtained the mean apparent source-time function duration of 1.0s, with its rupture towards the SW.

**Data and finite source analysis**

Accelerations taken at close distances from the source give an invaluable opportunity to study the strong motion in a wide rage of frequencies, including the static component of displacement. Here, the recordings due to the two largest events (foreshock and mainshock), obtained by the accelerometric stations within 25.0 km from the epicentral area (LOR, ZAR and AM2; Figure 1), were used to perform the analysis. Acceleration time histories were integrated two times following the procedure proposed by Iwan et al. (1985) and Zhu (2003). Due to its proximity from the source and the signal-to-noise ratio of accelerograms, an additional 3-pole Butterworth low-pass filtering with $f_c$=15.0hz was applied to the data (Boore and Bommer 2005). This procedure allows obtaining among others, the static component (permanent displacement) of the ground shaking. After the analysis, the near-field information was feasible to be recovered only from the LOR recordings. ZAR and AM2 stations presented strong path- and site-effects. In order to keep these effects from being transferred by the inversion into the slip distribution, the analysis was performed using the three components at LOR, for both earthquakes. The accelerometer of this station is oriented N30ºW (Alcalde, personal communication 2012). In order to obtain the NS and EW displacements, the horizontal components were clockwise rotated 30º before the inversion. Figure 2a and 2b shows the Vertical, NS and EW components of accelerations, due to the foreshock and mainshock respectively. The mainshock largest displacements (3.5cm) were found on the NS component of this station, with a permanent displacement of 1.2cm in the southern sense (Figure 4a, solid line). A vertical permanent displacement of 0.8cm was found on the vertical component in upward sense. The EW permanent displacements obtained have amplitudes below the noise level (Figure 4a, solid line).

**Table 1** Crustal velocity structure used in this study.

| # | $h$ | $v_p$ | $v_s$ | $\rho$ | $Q_p$ | $Q_s$ |
|---|-----|-------|-------|--------|-------|-------|
| 1 | 2.0 | 4.50 | 2.60 | 2.10 | 400 | 200 |
| 2 | 3.0 | 5.20 | 3.10 | 2.40 | 400 | 200 |
| 3 | 5.0 | 5.40 | 3.20 | 2.40 | 400 | 200 |
| 4 | 15.0 | 6.12 | 3.60 | 2.80 | 800 | 400 |
| 5 | 10.0 | 6.57 | 3.80 | 2.80 | 800 | 400 |
| 6 | $\infty$ | 7.76 | 4.50 | 2.90 | 800 | 400 |

# = layer number; h= layer thickness (km); $v_p$= P wave velocity (km/s); $v_s$= S wave velocity (km/s); $\rho$= mass density (g/cm$^3$); $Q_p$ = Quality factor for P waves; $Q_s$ = Quality factor for S waves. Layer 6 corresponds to the Halfspace.

To study the finite fault characteristics of the ruptures, a linear inversion of the observed displacement waveforms was performed to obtain the fault slip distribution. Due to the proximity of their hypocenters and the similarities among their acceleration waveforms, both analyzed events were assumed to occur over the same fault plane. For the inversion, rupture area was set as a rectangular fault of 4.0km x 4.0km in strike and dip directions respectively. This fault plane fits the rupture area of both events. The fault plane was discretized into 289 square subfaults of equal size (0.25km x 0.25km) embedded in a horizontally layered earth structure. Each subfault was modeled by a double couple point source located at its center. For each point source, the synthetic Green's functions at the recording site were computed using the discrete wave-number method described by Bouchon and Aki (1977) and Bouchon (1979). The crustal velocity structure in the vicinity of the source region was taken from Corchete and Chourak (2009) for the Murcia region (Table 1). The observed seismograms and the synthetic waveforms describe an over-determined system of linear equations of the form $\mathbf{Ax=b}$. Matrix $\mathbf{A}$ contains the synthetic seismograms with its respective time delay due to the rupture velocity, $\mathbf{b}$ is a vector with the observed records and $\mathbf{x}$ is the solution vector containing the dislocation weighting values for the slip of each subfault. A more detailed description of the inversion procedure can be found in Santoyo et al. (2005), which follows the methodology by Hartzell and Heaton (1983), and Mendoza and Hartzell (1989). For the inversion it was assumed a constant rupture velocity of 2.6 km/s, which is 0.85 times the S wave velocity at the depth of the fault plane. A constant velocity implies that the rupture propagates with circular fronts. The foreshock hypocenter was located at the centre of the fault plane and the mainshock hypocenter was located 1.0 km NE from the first one (Figures 3b and 4b). The corresponding hypocenter was assumed as the initiation point for the rupture. The location and depth of the mainshock hypocenter (37.727°N, 1.686°W, Z=4.5km) and the focal mechanism (Strike=240º, Dip=55º, Rake=45º) were obtained from Lopez-Comino et al. (2012). This mechanism was also used to define the spatial setting of the fault plane (Strike=240º, Dip=55º). Rise time at each subfault was set to 0.15s in order to maintain a smooth rupture over the entire fault. During the inversion procedure, it was assigned to all subfaults the same focal mechanism and rise time. The optimal smoothing parameters obtained for the foreshock and mainshock inversions were $\lambda_f$=3.0e-5 and $\lambda_m$=5.0e-4 respectively. The resulting slip distribution for the foreshock and the mainshock and the comparison between the synthetics and the observed displacements are shown in Figures 3 and 4 respectively. Figure 3a shows the comparison for the foreshock displacements. Here, observations are shown by solid lines and synthetics by dashed lines. Figures 3b and 3c show the resulting slip distributions. In a similar way, Figure 4a shows the comparison between the mainshock observed displacements and synthetics, and Figures 4b and 4c the resulting slip distributions. Here, a simple patch was obtained for the foreshock with a maximum slip of 6.0 cm and a static stress drop $\Delta\sigma \approx 65$ bar, assuming a circular fault. For the mainshock, two patches were obtained with the main one centred at its hypocenter. The maximum slip was on the north-eastern patch with a maximum of 23.0 cm. In this case, $\Delta\sigma \approx 45$ bar. Comparing the two obtained distributions, it appears that both events are complementary in their rupture areas, as it seems that no-overlapping exists between them. After inversion, the sense and amplitude of the static displacement agrees well with the observed data. The rupture distribution of mainshock is also consistent with the SW directivity obtained by Lopez-Comino et al. (2012), however in this case, rupture direction have a slightly down-dip propagation.

**Dynamic deformations.**

*Dynamic deformations using the Single Station method*

The 3D tensor components of the dynamic deformations at surface were first obtained from data by means of the single station method described in Gomberg (1997) and Singh et al. (1997). In this method, the displacement gradients used to calculate the dynamic deformations are obtained assuming the incidence of body waves through the recording site. In this way assuming S waves incidence, the horizontal particle motion can be written as

$$u_{SH} = A_{SH}e^{i(\omega t - k_h \cdot r + \theta_{SH})} \; ; \; u_{SVh} = A_{SVh}e^{i(\omega t - k_h \cdot r + \theta_{SVh})} \text{ and } u_{SVv} = A_{SVv}e^{i(\omega t - k_h \cdot r + \theta_{SVv})} \qquad (1)$$

where $u_{SH}$ is the SH motion in the transverse direction, $u_{SVh}$ and $u_{SVv}$ are the SV motions in the radial and vertical direction respectively; $A_{SH}$, $A_{SVh}$ and $A_{SVv}$ are the amplitudes in the transverse, radial and vertical directions and $\theta_{SH}$, $\theta_{SVh}$, $\theta_{SVv}$ are the phases of the incident wavefield at the surface in the measurement point. $\omega = 2\pi f$ is the angular frequency and $t=$ time. Here $\boldsymbol{k_h} \bullet \boldsymbol{r} = k_x x + k_y y$, where the magnitude of the horizontal wavenumber is $k_h = 2\pi/\lambda_h$ and $\lambda_h$ is the horizontal wavelength. This one can be expressed as $\lambda_h = TV_s/sin(\psi) = TV_h$, where $T=$period, $V_s$ is the S wave velocity, $V_h$ is the horizontal apparent velocity and $\psi$ is the angle of wave incidence with respect to the vertical.

The horizontal surface motion can be resolved in the three Cartesian directions as (e.g. Gomberg 1997)

$$u_j^x = A_j^x e^{i\theta_x} e^{i(\omega t - k_x x)} \text{ and } u_j^y = A_j^y e^{i\theta_y} e^{i(\omega t - k_y y)} \text{ ; } j=x,y,z \qquad (2)$$

The spatial derivatives of equations 2 are then taken in order to obtain the displacement gradients. The differentiation of the term $e^{-ik_x x}$ is equivalent to multiply by $-ik_x = -i2\pi/\lambda_x = -i(2\pi/\lambda_h)\sin(\phi)$ and the term $e^{-ik_y y}$ equivalent to multiply by $-ik_y = -i2\pi/\lambda_y = -i(2\pi/\lambda_h)\cos(\phi)$, where $\phi$=is the angle of azimuth of the incidence wavefield. The derivative with respect to time is equivalent to multiplication in the frequency domain by $i2\pi/T$. From this, the spatial gradients $G^s_{i,j}$ (s for single station) of equations 2, at the measuring site can be written as:

$$U_x = \frac{\partial u_1}{\partial x_1} = -\frac{\partial u_1}{\partial t}\frac{\sin\phi}{V_h} = -\frac{\partial u_1}{\partial t}\frac{\sin\phi\sin\psi}{V_s} \text{ ; } V_x = \frac{\partial u_2}{\partial x_1} = -\frac{\partial u_2}{\partial t}\frac{\sin\phi}{V_h} = -\frac{\partial u_2}{\partial t}\frac{\sin\phi\sin\psi}{V_s}$$

$$W_x = \frac{\partial u_3}{\partial x_1} = -\frac{\partial u_3}{\partial t}\frac{\sin\phi}{V_h} = -\frac{\partial u_3}{\partial t}\frac{\sin\phi\sin\psi}{V_s} \text{ ; } U_y = \frac{\partial u_1}{\partial x_2} = -\frac{\partial u_1}{\partial t}\frac{\cos\phi}{V_h} = -\frac{\partial u_1}{\partial t}\frac{\cos\phi\sin\psi}{V_s} \qquad (3)$$

$$V_y = \frac{\partial u_2}{\partial x_2} = -\frac{\partial u_2}{\partial t}\frac{\cos\phi}{V_h} = -\frac{\partial u_2}{\partial t}\frac{\cos\phi\sin\psi}{V_s} \text{ ; } W_y = \frac{\partial u_3}{\partial x_2} = -\frac{\partial u_3}{\partial t}\frac{\cos\phi}{V_h} = -\frac{\partial u_3}{\partial t}\frac{\cos\phi\sin\psi}{V_s}$$

Once obtained the components of $G^s_{i,j}$, these were used to derive the uniform strains by means of

$$\varepsilon_{i,j} = \frac{1}{2}\left(\frac{\partial u_i}{\partial x_j} + \frac{\partial u_j}{\partial x_i}\right) \qquad (4)$$

and the rigid body rotations by

$$\omega_{i,j} = \frac{1}{2}\left(\frac{\partial u_i}{\partial x_j} - \frac{\partial u_j}{\partial x_i}\right) \qquad (5)$$

where i=1,2,3; j=1,2,3; $u_1$=u, $u_2$=v, $u_3$=w; u, v and w are the displacements in the *x, y* and *z* directions at a given time. At the surface, due to the stress free boundary conditions, three components of $G^s_{i,j}$ are not independent: $\partial u_1/\partial x_3 = - \partial u_3/\partial x_1$ , $\partial u_2/\partial x_3 = -\partial u_3/\partial x_2$ and $\partial u_3/\partial x_3 = \eta \ (\partial u_2/\partial x_2 + \partial u_1/\partial x_1)$  and $\eta = -\lambda/ (\lambda+2\mu)$; $\lambda$ and $\mu$ are the Lamé parameters. From here, $\varepsilon_{1,3}=\varepsilon_{3,1} = \varepsilon_{2,3}= \varepsilon_{3,2}=0$.

*Dynamic deformations using a finite difference approach.*

In order to estimate the peak dynamic deformations in the Lorca urban area, the 3D synthetic displacement gradients tensor terms at surface were computed at the recording site. This modeling was based on the slip distributions obtained above and using a first-order finite difference scheme for the computation of the spatial derivatives of the ground motion.

In this way, the complete tensor of synthetic displacement gradients $G^f_{i,j}=\Delta u_i/\Delta x_j$ ($f$ for finite difference), where $\Delta u_i$, $i=x,y,z$, are the differences in the synthetic displacements in the three spatial directions and $\Delta x_j$, $j=1,2,3$  are the space increment in $x$, $y$ and $z$ directions respectively. The synthetic displacement time histories were computed by the discrete wave-number method (Bouchon 1979) within an area of 20 km x 20 km, centered on the recording site. The spacing among computing points was taken as $\Delta x_j = \lambda_{min}/100=V_s/100f_N$ , where $\Delta x_j=\Delta x_1=\Delta x_2=\Delta x_3$ , $V_s=$ minimum S wave velocity of the considered velocity structure, and $f_N$ =maximum analyzed frequency. The time increment for this analysis $\Delta t=1/f_s=1/2f_N$, where $f_s$=40.0Hz is the sampling frequency of the observed seismograms. Once obtained all the terms of $G^f_{i,j}$, the strain and rotation components were obtained applying equations 4 and 5. Cotton and Coutant (1997) tested this methodology with respect to the solution using analytic spatial derivatives, obtaining almost undistinguishable results between both techniques.

To check that the waveforms resulting from both methods used here agree in shape and amplitude, a quantitative comparison of the time histories was performed based on the misfit criterion:

$$m = 100 \frac{\sum_{k=1}^{N}\sqrt{\left[G^s_{i,j}(k\Delta t)-G^f_{i,j}(k\Delta t)\right]^2}}{A_{max}\sqrt{N}}$$

where $A_{max}$ is the maximum absolute amplitude of the time histories from the single station method, $\Delta t$ is the time increment and $N$ is the total number of time samples (e.g. Martinez-Garzón 2011). The computation of misfit was applied for a time window of 6.0 seconds after the first wave arrival.

**Results and discussion**

The wave incident angles $\phi$ and $\psi$ at surface used for the single station method, were obtained taking into account the subsoil velocity structure and the relative location of the hypocenter with respect to the recording site. For each time, all the terms of the displacement gradient tensor were computed using equations 3. The four non-vanishing terms of the strain tensor and the three rotational components at surface were obtained using equations 4 and 5. The results of applying this method to the velocity data at LOR are shown in Figures 5 and 6 with solid lines. Figure 5 shows the four tensor terms of the uniform strains and the three tensor terms of rotations obtained from the foreshock recordings. Figure 6 shows in a similar way, the non-vanishing terms due to the mainshock event. Here the maximum amplitudes on the displacement gradients are mostly obtained on the horizontal components. The $\varepsilon_{yy}$ term of uniform strains, systematically present the largest amplitudes. In the same way, rotation terms around the vertical axis are also larger than the ones around the horizontal axes. These results make sense due to the angle of incidence of the wave field, which in this case is highly vertical.

Figures 5 and 6 show with dashed lines, the synthetic deformations obtained from applying the finite difference method to the foreshock and the mainshock slip distributions. After the misfit analyses, the differences from both methods gave relatively low values (m<5%) except in the tilt terms (rotations with respect to the horizontal axes) where differences were higher (m<25%). Given this, the estimated dynamic deformations in the vicinity of the recording station were analyzed taking into account these differences.

Figures 7 and 8 show the resulting distribution of surface peak-dynamic strains and rotations due to the mainshock. In these Figures, Lorca town contour is shown in solid brown. The aftershock locations relocated by López-Comino et al. (2012) are shown with open dots. The surface projection of the fault plane with respect to the urban area is shown with a black rectangle. The relative location between the fault plane and aftershocks suggest that they are complementary to the foreshock and the main ruptures.

Figures 7a to 7f shows the negative and positive peak dynamic values of the strain tensor terms $\varepsilon_{1,1}$, $\varepsilon_{2,2}$ respectively. Figures 7g and 7h shows the negative and positive values of $\varepsilon_{3,3}$ $\varepsilon_{1,2}$ respectively. In these figures, units are in strain. Colour scale is shown at the right hand side for each Figure, where cold colours show the negative values and the warm colours the positive range. Figures 8a to 8d show the clockwise and counter-clockwise peak dynamic tilts in the $x$ and $y$ directions respectively. Figures 8e and 8f show the clockwise and counter-clockwise peak dynamic torsions (rotations along the vertical axis $z$). Here, units are in radians and the colour scale is shown at the right hand side of each map. Cold colours show the clockwise values and warm colours the counter-clockwise range.

From the maps showing the distribution of strain terms $\varepsilon_{2,2}$, $\varepsilon_{3,3}$ and $\varepsilon_{1,2}$, it can be observed that the zones of maximum peak dynamic strains fall on the northern and southern sides from the ruptured fault. The southern zone appears to be coincident with the location of the urban area. Absolute values in these cases range in the urban area between $3.5\times10^{-5}$ and $1.0\times10^{-4}$ strain. The maximum peak dynamic strains corresponding to the term $\varepsilon_{1,1}$ seems to occur on the south-western and north-eastern sides of the town. The order of magnitude of these strains, suggests that surface deformations might have contributed to the damage of buried lifelines (e.g. Singh et al. 1997; Trifunac et al. 1996; Pineda-Porras and Najafi 2010). In any case, no evident damages were observed on large infrastructures like tunnels or bridges in the zone.

On the other hand, distribution of tilts around the $y$ axis seems to occur mainly in the eastern and western sides of the rupture; here the urban area appears to be in a low value zone. On the contrary, peak dynamic tilts around $x$ axis and peak dynamic torsions could have their maximums near the town zone, however, the absolute values on tilts for this case are not reliable. Maximum peak torsions in some parts of the urban area could reach values of $3.0\times10^{-4}$ radians. In the same sense, the counter-clockwise and clockwise rotations are in the urban area of the same order of magnitude. Due to this, a direct relation between these results and the observed rotations is difficult to achieve. This task could have additional difficulties because translational motions can produce some amount of the observed permanent rotations (e.g. Teisseyre et al. 2003; Hinzen 2012).

Relative amplitudes of deformations however, can give some clues about the distribution of them in the near region from the fault at surface. The disagreement found on tilts could be possibly attributed to 2D or 3D propagation effects. For example the effects of topography on local wave amplification could be, under some circumstances, much larger on the horizontal directions (e.g. Sanchez-Sesma and Campillo (1991), Paolucci, 2002). On the other hand, better adjustments on these terms could be obtained performing a non-linear kinematic inversion. Given the size of the studied earthquakes, a non-linear analysis was not the first-election inversion procedure, because this could introduce additional complexities on slips that may not be real.

**CONCLUSIONS**

Slip distributions obtained from kinematic inversion suggest that the two analyzed events, i.e. the Mw4.6 foreshock and the Mw5.2 mainshock, may have complementary ruptures as they are not overlapped. A more detailed analysis by stress transfer computations over the fault plane could be realized in order to study the possible stress triggering of the mainshock due to the previous event. In any case, this kind of analysis was outside the scope of this work. Slip distributions show for the first event, a single ruptured patch with a maximum slip of 6.0 cm. For the second event two main patches were obtained with a maximum slip of 23.0 cm.

Peak dynamic strains at the Lorca accelerometric station could reach values of the order of $1.2 \times 10^{-4}$ strain, especially on the $\varepsilon_{2,2}$ term of the dynamic tensor, and values of $2.0 \times 10^{-4}$ radians in the $\omega_3$ term. The order of magnitude of these values suggests that surface strains could contribute to produce damage to some shallow buried lifelines. On the other hand, the obtained absolute values on torsions could contribute to increase the rotational effects on structures.

Maximum peak dynamic strain terms $\varepsilon_{2,2}$, $\varepsilon_{3,3}$ and $\varepsilon_{1,2}$, seems that occurred in the same zone of the urban area of Lorca. The same situation occurs with the rotational terms $\omega_1$, $\omega_3$ where maximum peak dynamic rotations appear to occur also near the urban area.

Relative values of strains and rotations can give clues about the behaviour of dynamic strains and rotations experimented in the vicinity of the Lorca town during the Mw5.2 mainshock and the Mw4.6 foreshock. However its effects on the observed structural damages should be more investigated.

**Acknowledgments**

Accelerometric data were supplied by the Instituto Geografico Nacional de España (IGN). I wish to thank Resurreción Antón and Juan Manuel Alcalde from IGN for their kind assistance in downloading the seismic data. Thanks to José Morales from Instituto Andaluz de Geofísica for providing the relocated seismicity. Fruitful discussions with Jorge M. Gaspar-Escribano, Miguel Herraiz and the constructive comments of two anonymous reviewers helped to improve this work. Ligia E. Quiroz and Patricia Martinez-Garzón helped in part of the analysis of the theoretical deformations. DEGTRA-A4 program was used to process the accelerometric data. Axitra program by O. Coutant was used for the computation of Green's Functions. This work was done while the author M.A.S. was under the Subprogram Ramón y Cajal contract, of the Ministerio de Economía y Competitividad, Spain, supported by the European Social Fund.

**FIGURE CAPTIONS**

**Fig. 1** Location map and tectonic setting of the studied area. Solid lines indicate the trace of main faults. Solid triangles show the location of the accelerometric stations: LOR=Lorca; ZAR= Zarcilla de Ramos; AM2= Alhama de Murcia. Solid diamonds show the location of nearby towns. White star show the 11/5/2011 (Mw5.2) mainshock location. Gray rectangle shows the area for computations of surface dynamic deformations. Inset: General location map of the study area in Spain.

**Fig. 2** Vertical, NS and EW acceleration recordings at station LOR, for the Mw4.7 foreshock (2a) and the Mw5.2 mainshock (2b)

**Fig. 3** Results from the linear waveform inversion for the slip distribution due to the 11/5/2011 foreshock. **a.** Observed (solid) and synthetic (dashed) displacements at LOR in the NS (bottom), EW (middle) and vertical (top) directions. Displacement amplitudes are in cm. **b.** Slip distribution from kinematic inversion. Fault plane is viewed from the NW. Black star shows the location of the epicentre and the initiation point for the rupture. Epicentre is at 4.5 km depth. **c.** Perspective view of the rupture amplitudes

**Fig. 4** Results from the linear waveform inversion for the slip distribution due to the 11/5/2011 mainshock. Notes are the same as in Figure 3

**Fig 5** Foreshock dynamic strains and rotations at LOR station. Uniform strains $\varepsilon_{11}$, $\varepsilon_{22}$, $\varepsilon_{33}$ and $\varepsilon_{12}$ are shown at top and $\omega_1$, $\omega_2$ and $\omega_3$, at bottom. Solid lines show the time histories from the single station method and dashed lines show histories from the finite difference modeling. Units for uniform strains are in strain, and radians for rigid body rotations

**Fig 6** Mainshock dynamic strains and rotations at LOR station. Uniform strains $\varepsilon_{11}$, $\varepsilon_{22}$, $\varepsilon_{33}$ and $\varepsilon_{12}$ are shown at the top of figure and $\omega_1$, $\omega_2$ and $\omega_3$, at the bottom. Notes are the same as Figure 5

**Fig 7** Mainshock peak dynamic strain maps at the surface. **a** and **b:** negative and positive peak dynamic values of the strain tensor term $\varepsilon_{1,1}$ respectively. **c** and **d:** peak values of $\varepsilon_{2,2}$. **e** and **f:** values of $\varepsilon_{3,3}$. **g** and **h**: values of $\varepsilon_{1,2}$. In all cases units are in strain. Colour scale is shown at the right hand side of each map, where cold colours show the negative values and the warm colours the positive range

**Fig 8** Mainshock peak dynamic rotation maps at surface. **a** and **b:** Clockwise and counter-clockwise peak dynamic tilts (rotations with respect to the horizontal axis) in $x$ direction. **c** and **d:** tilts in $y$ direction. **e** and **f:** torsions in $z$ direction (rotations along the vertical axis $z$). Units in all cases are in radian. Colour scale is shown at the right hand side of each map. Cold colours show the clockwise values and warm colours the counter-clockwise range

**FIGURES**

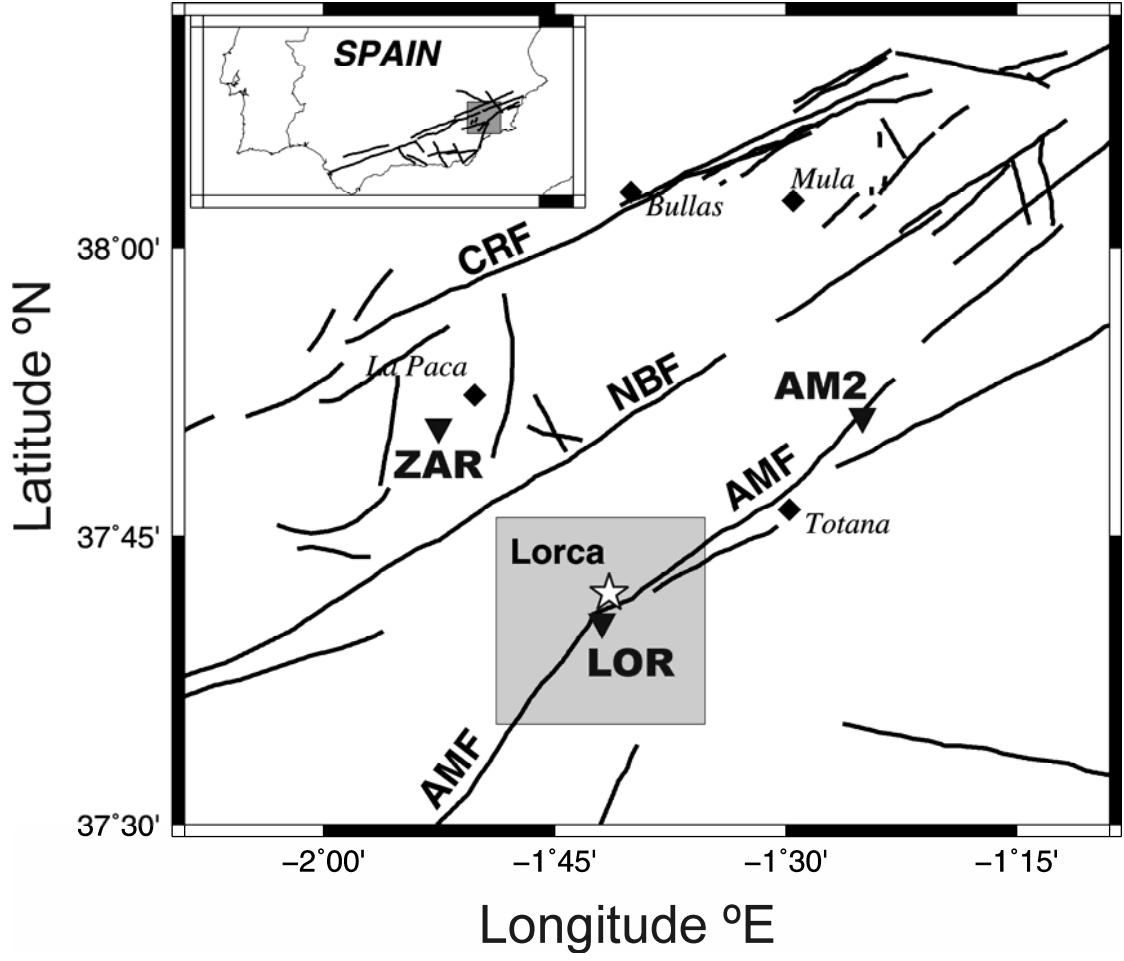

Figure 1

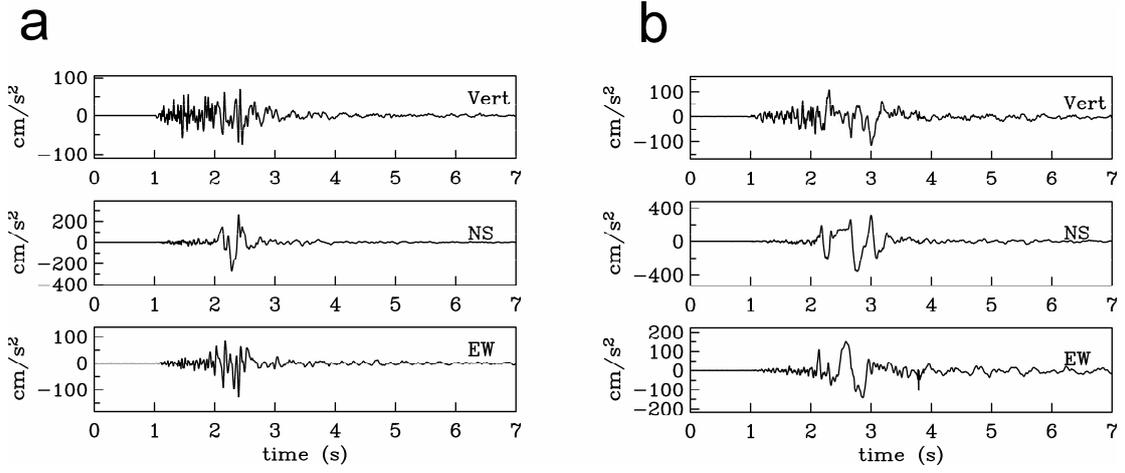

**Figure 2**

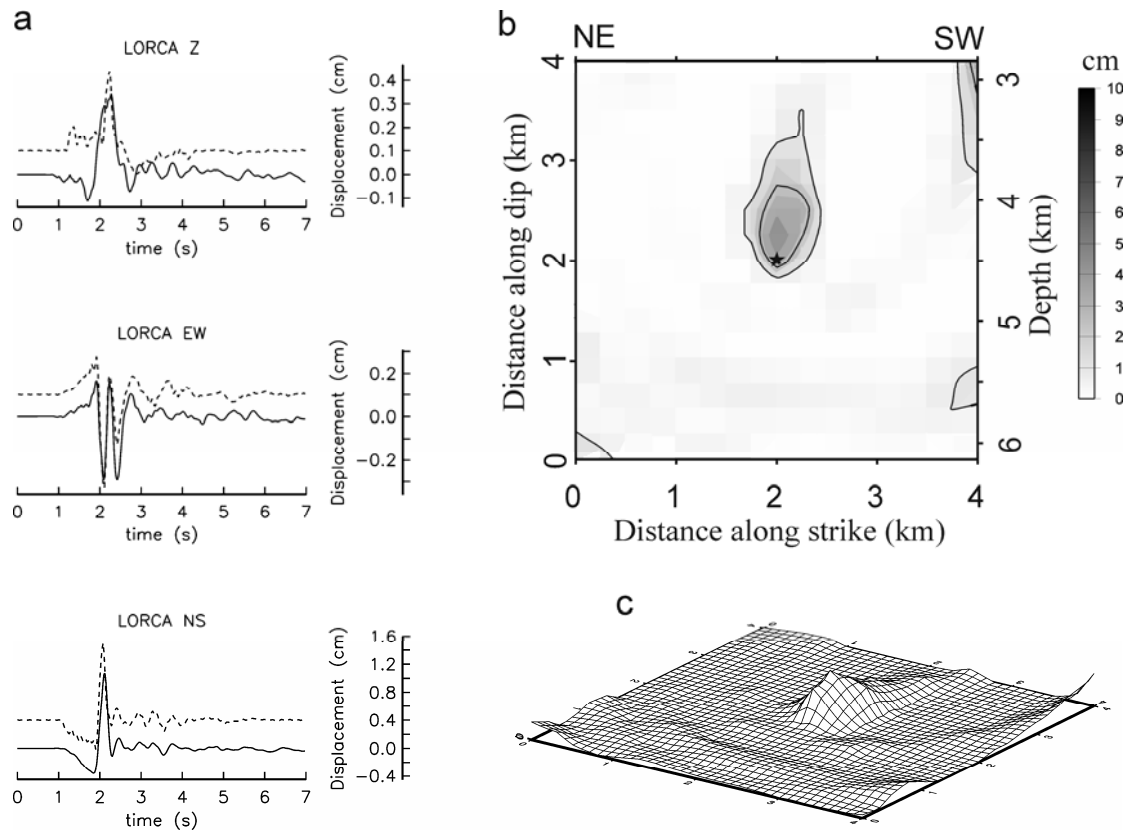

**Figure 3**

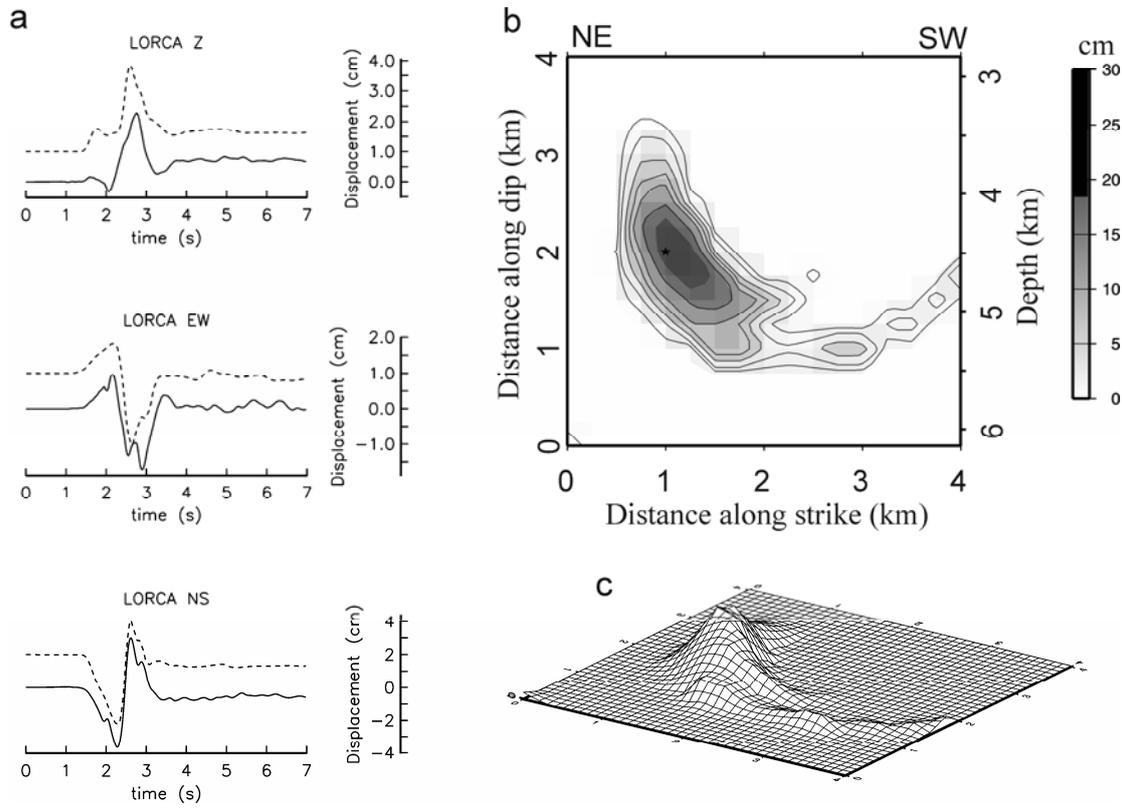

**Figure 4**

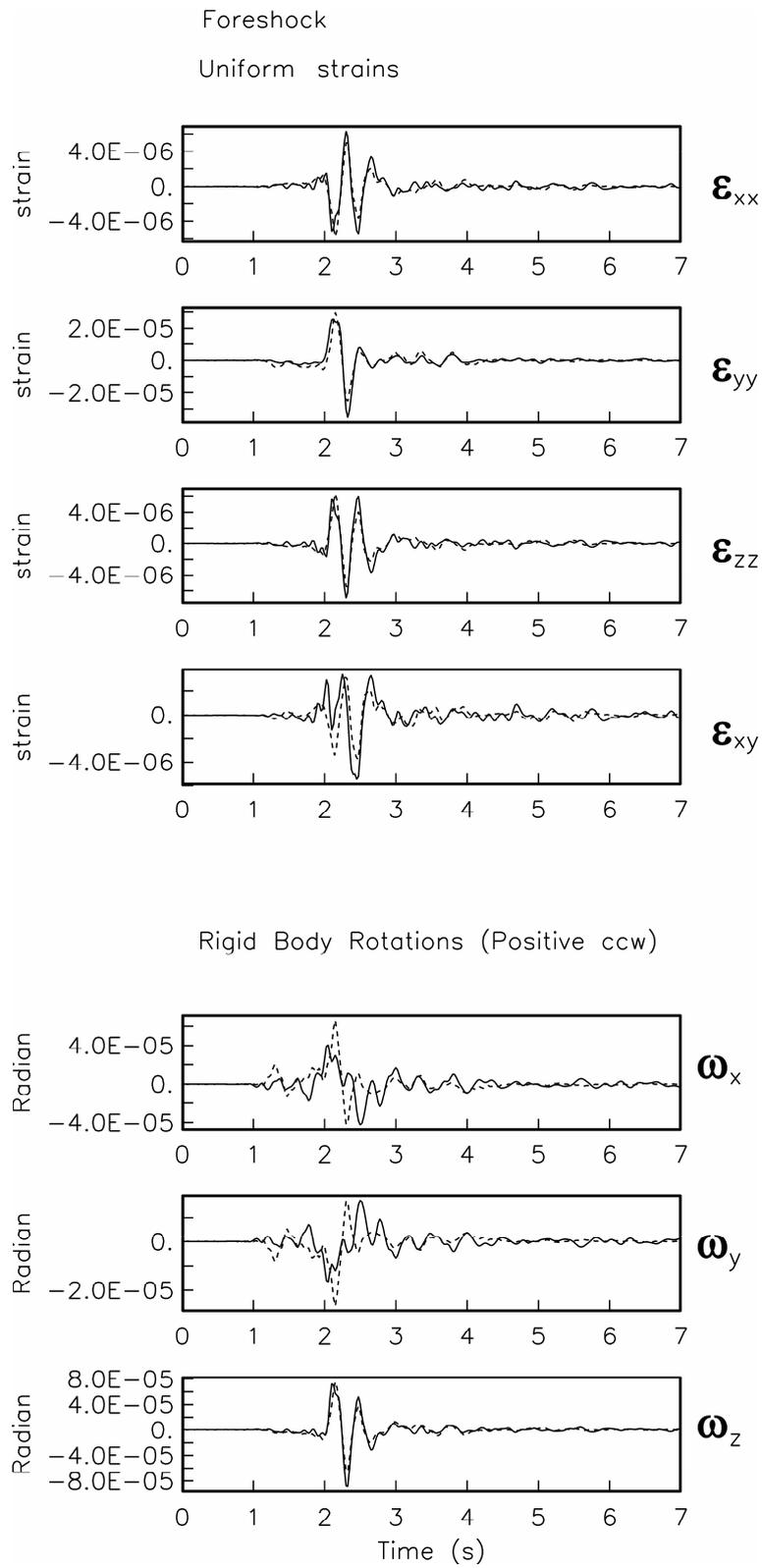

Figure 5

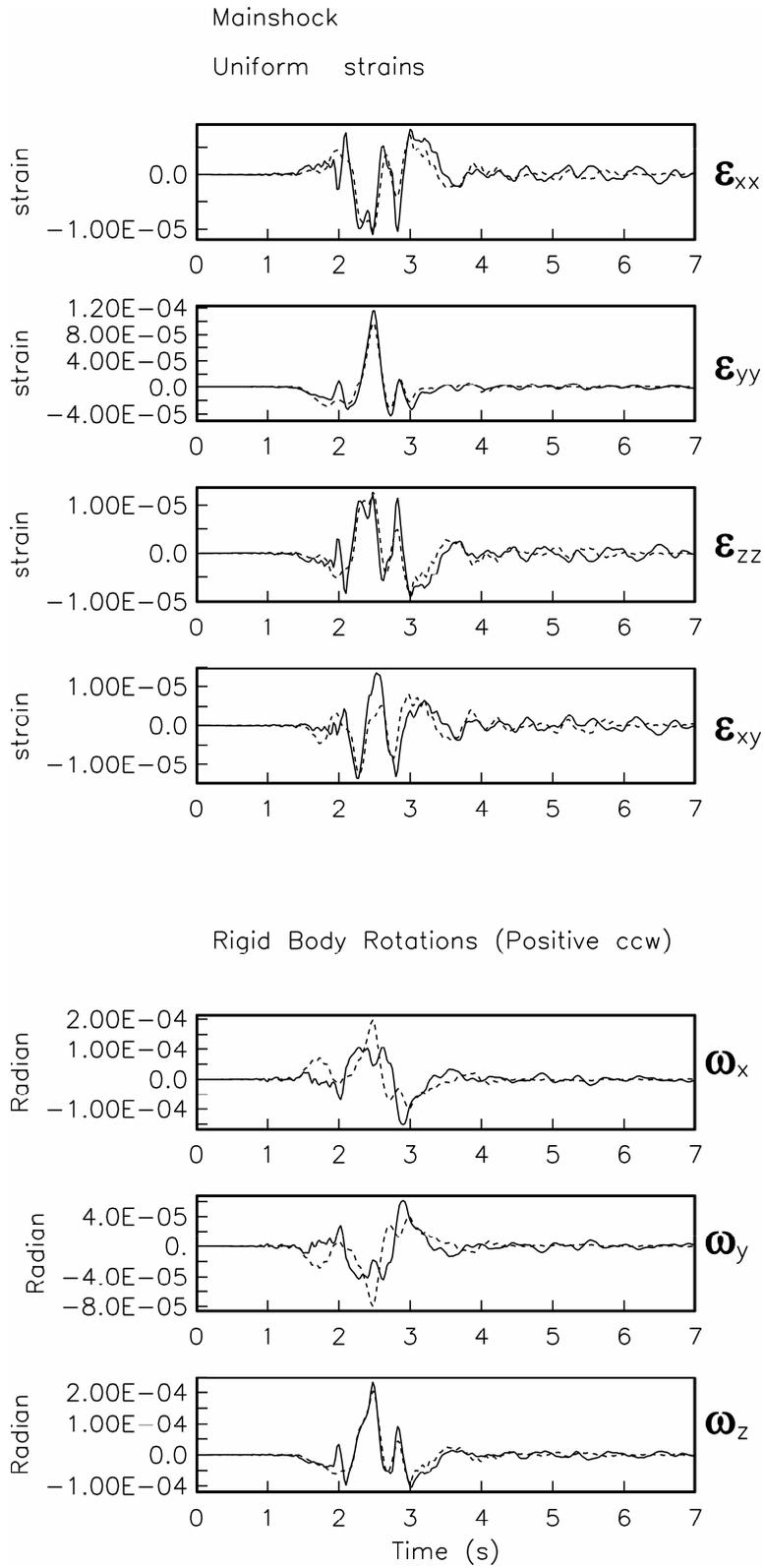

Figure 6

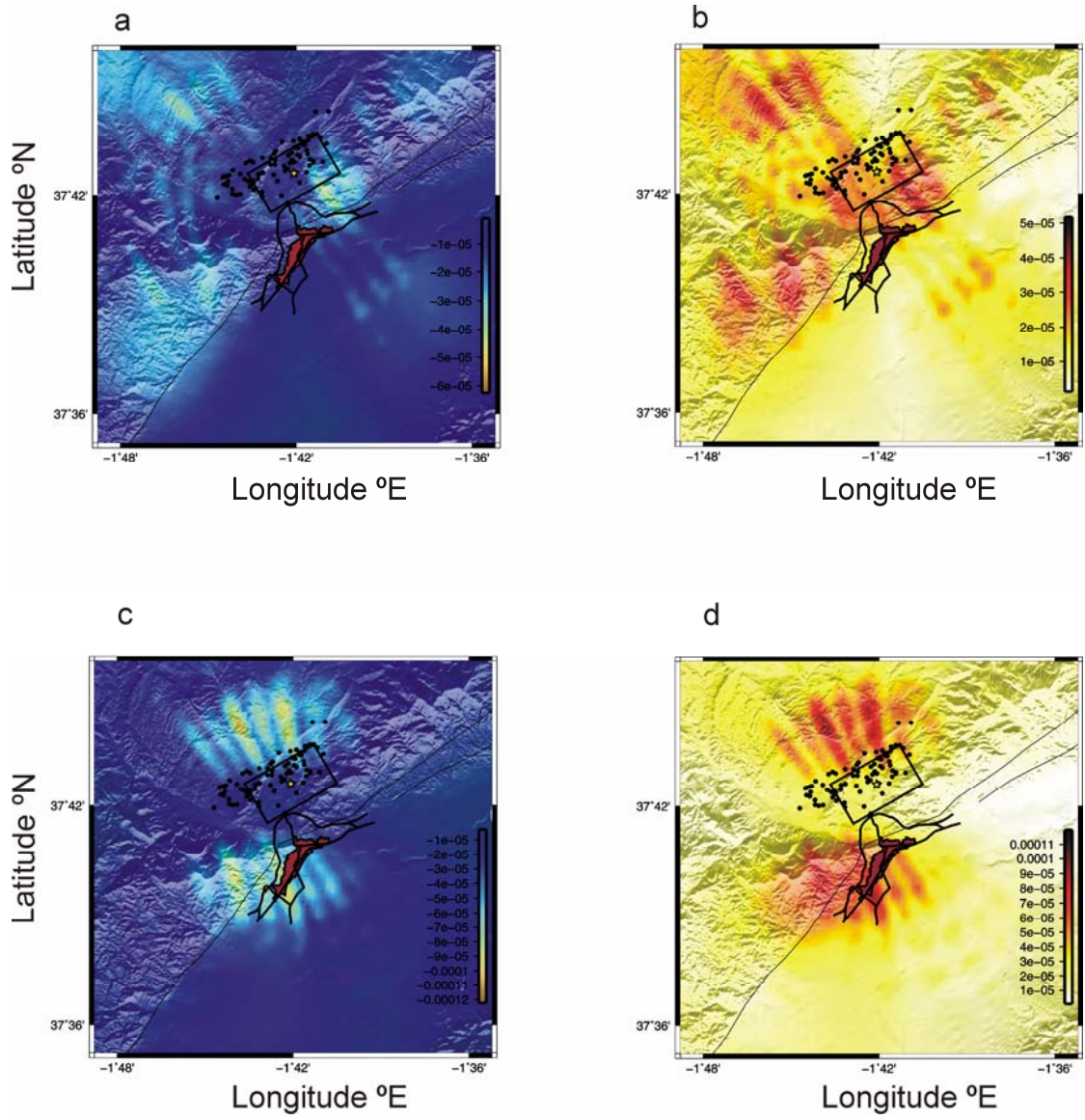

**Figure 7**

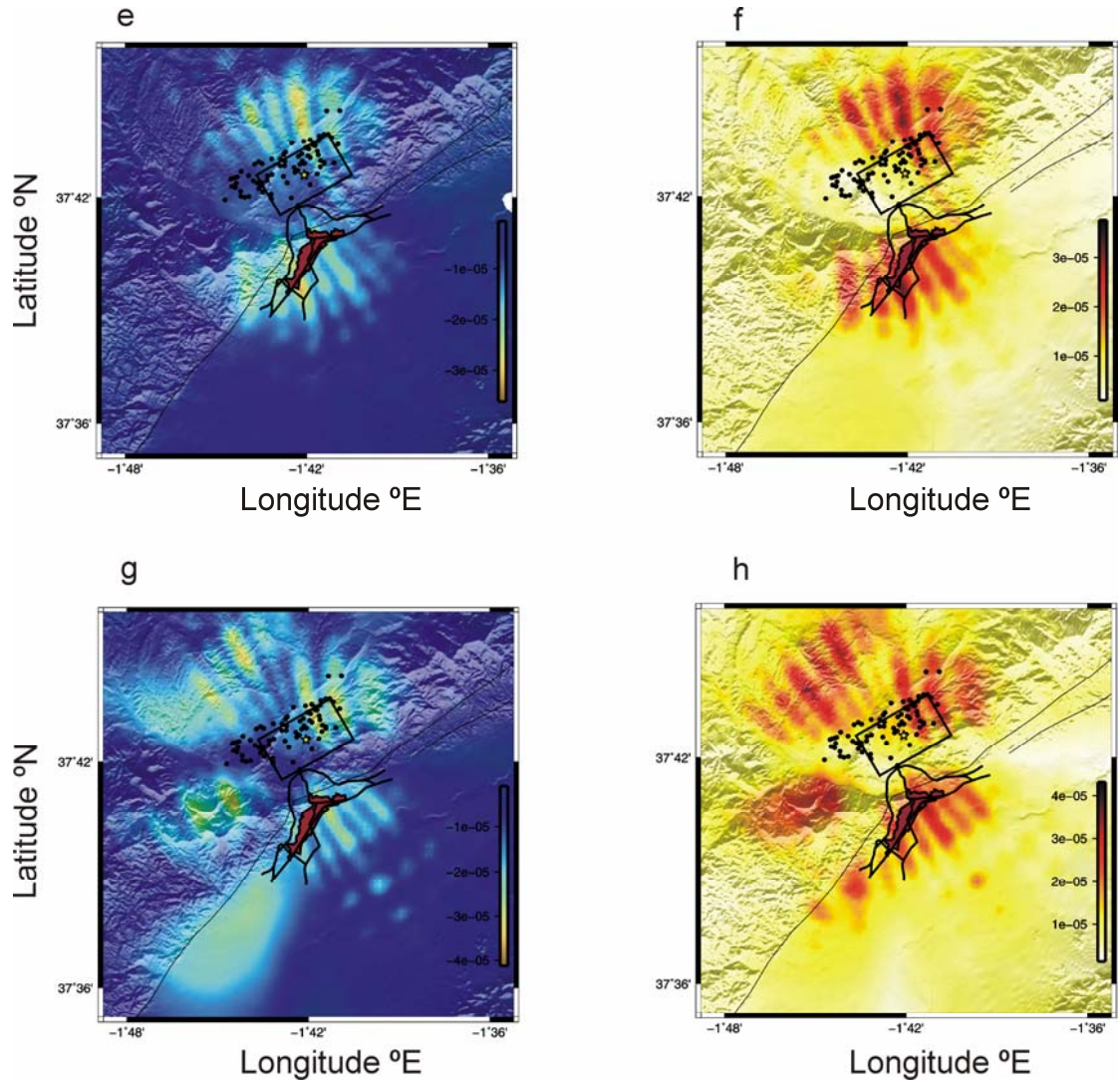

**Figure 7 (cont)**

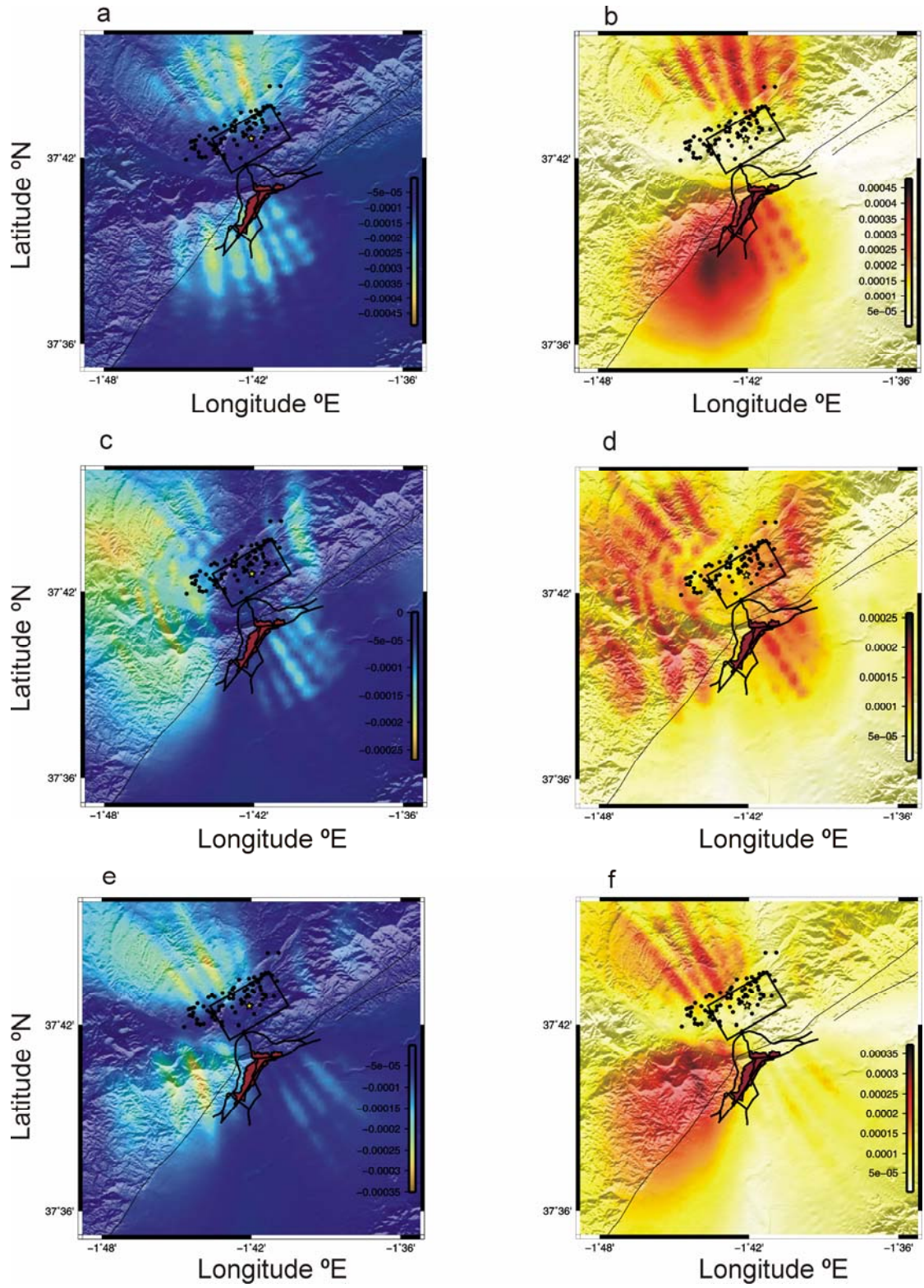

**Figure 8**